\documentclass[preprint,showpacs,preprintnumbers,amsmath,amssymb]{revtex4}

\usepackage{graphicx}

\begin{document}

\title{
Uniqueness of collinear solutions  
for the relativistic three-body problem 
}
\author{Kei Yamada}
%\email{}
\author{Hideki Asada} 
%\email{asada@phys.hirosaki-u.ac.jp}
\affiliation{
Faculty of Science and Technology, Hirosaki University,
Hirosaki 036-8561, Japan} 

\date{\today}

\begin{abstract}
Continuing work initiated in an earlier publication 
[Yamada, Asada, Phys. Rev. D {\bf 82}, 104019 (2010)], 
we investigate collinear solutions to 
the general relativistic three-body problem. 
We prove the uniqueness of the configuration 
for given system parameters (the masses and 
the end-to-end length). 
First, we show that the equation determining the distance ratio 
among the three masses, 
which has been obtained as a seventh-order polynomial 
in the previous paper, 
has at most three positive roots, 
which apparently provide three cases of the distance ratio. 
It is found, however, that, even for such cases, there exists 
one physically reasonable root and only one, 
because the remaining two positive roots do not satisfy 
the slow motion assumption in the post-Newtonian approximation 
and are thus discarded. 
This means that, especially for the restricted three-body problem, 
exactly three positions of a third body are true even 
at the post-Newtonian order. 
They are relativistic counterparts of 
the Newtonian Lagrange points $L_1$, $L_2$ and $L_3$. 
We show also that, for the same masses and full length, 
the angular velocity of the post-Newtonian 
collinear configuration is smaller than that for the Newtonian case.  
Provided that the masses and angular rate are fixed, 
the relativistic end-to-end length is shorter than the Newtonian one. 
\end{abstract}


\maketitle

\section{Introduction}
%\noindent \emph{Introduction.--- } 
The three-body problem in Newtonian gravity belongs 
among classical problems in astronomy and physics 
(e.g, \cite{Danby,Goldstein}). 
In 1765, Euler found a collinear solution for 
the restricted three-body problem, 
where one of three bodies is a test mass. 
Soon after, his solution was extended for 
a general three-body problem by Lagrange, 
who also found an equilateral triangle solution 
in 1772. 
Now, the solutions for the restricted three-body problem 
are called Lagrange points $L_1, L_2, L_3, L_4$ and $L_5$, 
which are described 
in textbooks of classical mechanics \cite{Goldstein}. 
SOHO (Solar and Heliospheric Observatory) and 
WMAP (Wilkinson Microwave Anisotropy Probe) 
launched by NASA are in operation 
at the Sun-Earth $L_1$ and $L_2$, respectively. 
LISA (Laser Interferometer Space Antenna) pathfinder 
is planned to go to $L_1$. 
Lagrange points have recently attracted renewed interests 
for relativistic astrophysics \cite{THA,Asada,SM,Schnittman}, 
where they have discussed the gravitational radiation 
reaction on $L_4$ and $L_5$ analytically \cite{Asada}
and by numerical methods \cite{THA,SM,Schnittman}. 

As a pioneering work, Nordtvedt pointed out that 
the location of the triangular points is very sensitive 
to the ratio of the gravitational mass to the inertial one 
\cite{Nordtvedt}. 
Along this course, it is interesting as a gravity experiment 
to discuss the three-body coupling terms at the post-Newtonian 
order, 
because some of the terms are proportional to a product of 
three masses as $M_1 \times M_2 \times M_3$. 
Such a term appears only for relativistic three (or more) 
body systems: 
For a relativistic binary with two masses $M_1$ and $M_2$, 
there exist $M_1^2 M_2$ and $M_1 M_2^2$ without 
such a three-mass product. 
For a Newtonian three-body system, we have 
only the two-body coupling terms proportional to 
$M_1 M_2$, $M_2 M_3$ or $M_3 M_1$. 

The relativistic perihelion advance of Mercury 
is detected only after much larger shifts due to 
Newtonian perturbations by other planets such as 
the Venus and Jupiter are taken into account 
in the astrometric data analysis. 
In this sense, effects by the three body coupling 
are worthy to investigate. 
Nevertheless, most of post-Newtonian works have focused on 
either compact binaries 
because of our interest in gravitational waves astronomy 
or 
N-body equation of motion (and coordinate systems) 
in the weak field such as the solar system (e.g. \cite{Brumberg}). 
Actually, future space astrometric missions 
such as Gaia 
\cite{GAIA,JASMINE}
require a general relativistic modeling of 
the solar system within the accuracy of a micro arc-second 
\cite{Klioner}. 
Furthermore, a binary plus a third body have been discussed 
also for perturbations of gravitational waves induced by the third body 
\cite{ICTN,Wardell,CDHL,GMH}. 

After efforts to find a general solution, 
Poincare proved that 
it is impossible to describe all the solutions 
to the three-body problem even for the $1/r$ potential. 
Namely, we cannot analytically obtain all the solutions. 
Nevertheless, the number of new solutions is increasing \cite{Marchal}.   
Therefore, the three-body problem still remains an open issue 
even for Newton gravity. 

The theory of general relativity is currently 
the most successful gravitational theory describing 
the nature of space and time. 
Hence, it is important to take account of general relativistic effects 
on three-body configurations. 
The figure-eight configuration that was found decades ago 
\cite{Moore,CM} has been recently studied 
at the first post-Newtonian \cite{ICA} 
and also the second post-Newtonian orders \cite{LN}.
According to their numerical investigations, 
the solution remains true with a slight change 
in the figure-eight shape because of relativistic effects.

On the other hand, the post-Newtonian collinear configuration 
obtained in the previous paper \cite{YA} may offer 
a useful toy model for relativistic three-body interactions, 
because it is tractable by hand without numerical simulations. 
This solution is a relativistic extension of 
{\it Euler's collinear one}, 
where three bodies move around the common center of 
mass with the same orbital period and always line up. 

In fact, their formulation leads to a seventh-order equation 
determining the distance ratio among masses \cite{YA}. 
Here, it should be noted that only positive roots are acceptable, 
because the distance ratio must be positive. 
Properties of the master equation have not been known yet. 
How many positive roots for it are there? 
The main purpose of this paper is to {\it analytically} investigate 
the number of the positive roots. 
In particular, we shall prove the uniqueness of the configuration 
for given system parameters (the masses and the end-to-end length). 

This paper is organized as follows. 
In section II, we briefly summarize formulations 
for collinear solutions at the Newtonian and 
post-Newtonian orders. 
We discuss positive roots for the seventh-order equation 
for determining the distance ratio in section III. 
In section IV, we show the uniqueness of the configuration 
for given system parameters (the masses and the end-to-end length). 
We also compare the angular velocity of the post-Newtonian 
collinear configuration with that for the Newtonian one. 
Section V is devoted to the conclusion. 
We provide some detailed calculations regarding the angular 
velocity of collinear configurations in the Appendix. 
 
Throughout this paper, we take the units of $G=c=1$.

\section{Equation for the distance ratio among three masses} 
%\noindent \emph{Newtonian Euler's collinear solution.---} 
Let us begin by summarizing the derivation of 
the Euler's collinear solution for the circular three-body problem 
in Newton gravity. 
We consider Euler's solution, for which  
each mass moves around their common center 
of mass denoted as $\mbox{\boldmath $X$}_G$ 
with a constant angular velocity $\omega$. 
Hence, it is convenient to use the corotating frame 
with the same angular velocity $\omega$. 
We choose an orbital plane normal to the total angular momentum 
as the $x-y$ plane in such a corotating frame. 
We locate all the three bodies on a single line, 
along which we take the $x$-coordinate. 
The location of each mass $M_I$ $(I=1, 2, 3)$ is written as
$\mbox{\boldmath $X$}_I \equiv (x_I, 0)$. 
Without loss of generality, 
we assume $x_3 < x_2 < x_1$. 
Let $R_I$ define the relative position 
of each mass with respective to the center of mass 
$\mbox{\boldmath $X$}_G \equiv (x_G, 0)$, 
namely $R_I \equiv x_I - x_G$ 
($R_I \neq |\mbox{\boldmath $X$}_I|$ unless $x_G = 0$). 
We choose $x=0$ between $M_1$ and $M_3$. 
We thus have $R_3 < R_2 < R_1$, $R_3 < 0$ and $R_1 > 0$. 

It is convenient to define a ratio as 
$R_{23} / R_{12}= z$, which is an important variable 
in the following formulation. 
Then we have $R_{13} = (1+z) R_{12}$. 
The equation of motion becomes 
\begin{eqnarray}
R_1 \omega^2 &=& \frac{M_2}{R_{12}^2} + \frac{M_3}{R_{13}^2} , 
\label{EOM-M1-N}
\\
R_2 \omega^2 &=& -\frac{M_1}{R_{12}^2} + \frac{M_3}{R_{23}^2} , 
\label{EOM-M2-N}
\\
R_3 \omega^2 &=& -\frac{M_1}{R_{13}^2} - \frac{M_2}{R_{23}^2} , 
\label{EOM-M3-N} 
\end{eqnarray}
where we define 
\begin{eqnarray}
\mbox{\boldmath $R$}_{IJ} &\equiv& 
\mbox{\boldmath $X$}_{I}-\mbox{\boldmath $X$}_{J} , 
\\
R_{IJ} &\equiv& |\mbox{\boldmath $R$}_{IJ}| . 
\end{eqnarray}

First, we subtract Eq. ($\ref{EOM-M2-N}$) from Eq. ($\ref{EOM-M1-N}$) 
and Eq. ($\ref{EOM-M3-N}$) from Eq. ($\ref{EOM-M2-N}$) 
and use 
$R_{12} \equiv |\mbox{\boldmath $X$}_1 - \mbox{\boldmath $X$}_2|$ 
and 
$R_{23} \equiv |\mbox{\boldmath $X$}_2 - \mbox{\boldmath $X$}_3|$. 
Such a subtraction procedure will be useful 
also at the post-Newtonian order, 
because we can avoid directly using  
the post-Newtonian center of mass \cite{MTW,LL}. 
Next, we compute a ratio between them to delete $\omega^2$. 
Hence a fifth-order equation is obtained as 
\begin{equation}
(M_1+M_2) z^5 + (3M_1+2M_2)z^4 + (3M_1+M_2)z^3 
- (M_2+3M_3)z^2 - (2M_2+3M_3)z - (M_2+M_3) = 0. 
\label{5th}
\end{equation}
Now we have a condition as $z>0$. 
Descartes' rule of signs (e.g., \cite{Waerden}) 
states that the number of positive roots either equals 
that of sign changes in coefficients of a polynomial or 
less than it by a multiple of two. 
According to this rule, 
Eq. ($\ref{5th}$) has only the positive root $z>0$, 
though such a fifth-order equation cannot be solved 
in algebraic manners as shown by Galois (e.g., \cite{Waerden}). 
After obtaining $z$, one can substitute it into a difference, 
for instance between Eqs. ($\ref{EOM-M1-N}$) and ($\ref{EOM-M3-N}$). 
Hence we get $\omega$. 
 
In order to include the dominant part of general relativistic effects, 
we take account of the terms at the first post-Newtonian order. 
Namely, the massive bodies obey  
the Einstein-Infeld-Hoffman (EIH) equation of motion as \cite{MTW,LL} 
\begin{eqnarray}
\frac{d \mbox{\boldmath $v$}_K}{dt} 
&=& \sum_{A \neq K} \mbox{\boldmath $R$}_{AK} 
\frac{M_A}{R_{AK}^3} 
\left[
1 - 4 \sum_{B \neq K} \frac{M_B}{R_{BK}} 
- \sum_{C \neq A} \frac{M_C}{R_{CA}} 
\left( 1 - 
\frac{\mbox{\boldmath $R$}_{AK} \cdot \mbox{\boldmath $R$}_{CA}}
{2R_{CA}^2} \right) 
\right.
\nonumber\\
&&
\left. 
~~~~~~~~~~~~~~~~~~~~~
+ v_K^2 + 2v_A^2 - 4\mbox{\boldmath $v$}_A \cdot \mbox{\boldmath $v$}_K 
- \frac32 \left( 
\mbox{\boldmath $v$}_A \cdot \mbox{\boldmath $n$}_{AK} \right)^2 
\right]
\nonumber\\
&&
- \sum_{A \neq K} (\mbox{\boldmath $v$}_A - \mbox{\boldmath $v$}_K) 
\frac{M_A \mbox{\boldmath $n$}_{AK} \cdot 
(3 \mbox{\boldmath $v$}_A - 4 \mbox{\boldmath $v$}_K)}{R_{AK}^2} 
\nonumber\\
&&
+ \frac72 \sum_{A \neq K} \sum_{C \neq A} 
\mbox{\boldmath $R$}_{CA} 
\frac{M_A M_C}{R_{AK} R_{CA}^3} , 
\label{EIH-EOM}
\end{eqnarray}
where $\mbox{\boldmath $v$}_I$ denotes the velocity of each mass 
in an inertial frame 
and we define 
\begin{eqnarray}
\mbox{\boldmath $n$}_{IJ}&\equiv&
\frac{\mbox{\boldmath $R$}_{IJ}}{R_{IJ}} , 
\end{eqnarray}
and we assume the slow motion ($|\mbox{\boldmath $v$}_I| \ll c$). 

We obtain a lengthy form of the equation of motion for each body. 
By subtracting the post-Newtonian equation of motion 
for $M_3$ from that for $M_1$ for instance, 
we obtain the equation as \cite{YA}
\begin{equation}
R_{13} \omega^2 = F_N + F_M + F_V \omega^2 , 
\label{EOM-PN}
\end{equation}
where we denote $a \equiv R_{13}$ and  
the Newtonian term $F_N$ 
and the post-Newtonian parts $F_M$ (dependent on the masses only) 
and $F_V$ (velocity-dependent part divided by $\omega^2$) 
are defined as 
\begin{eqnarray}
F_N &=& \frac M{a^2z^2} 
\Biggl[
 (\nu_1 + \nu_3) z^2 + (1 - \nu_1 - \nu_3) (1 + z^2) (1 + z)^2
\Biggr],
\label{FN}
\\
F_M &=& - \frac {M^2}{a^3 z^3} 
\Biggl[
 (4 - 4\nu_1 + \nu_3) (1 - \nu_1 - \nu_3)
\notag \\
&& \mspace{65mu}
+ (12 - 7\nu_1 + 3\nu_3) (1 - \nu_1 - \nu_3) z
\notag \\
&& \mspace{65mu}
+ (12 - \nu_1 + \nu_3) (1 - \nu_1 - \nu_3) z^2
\notag \\
&& \mspace{65mu}
+ (8 - 7\nu_1 - 7\nu_3 + 8\nu_1\nu_3 + 3\nu_1^2 + 3\nu_3^2) z^3
\notag\\
&& \mspace{65mu}
+ (12 + \nu_1 - \nu_3) (1 - \nu_1 - \nu_3) z^4
\notag \\
&& \mspace{65mu}
+ (12 + 3\nu_1 - 7\nu_3) (1 - \nu_1 - \nu_3) z^5
\notag \\
&& \mspace{65mu}
+ (4 +  \nu_1 - 4\nu_3) (1 - \nu_1 - \nu_3) z^6 
\Biggr],
\label{FM}
\\
F_V &=& \frac M{(1 + z)^2 z^2}
\Biggl[
- \nu_1^2 (1 - \nu_1 - \nu_3)
\notag \\
&& \mspace{91mu}
 - 2\nu_1 (1 + \nu_1 - \nu_3) (1 - \nu_1 - \nu_3) z
\notag \\
&& \mspace{91mu}
+ (2 - 2\nu_1 + \nu_3 + 6\nu_1\nu_3 - 3\nu_3^2 
+ \nu_1^3 - 3\nu_1^2\nu_3 - 3\nu_1\nu_3^2 + \nu_3^3) z^2
\notag \\
&& \mspace{91mu}
+ 2(2 - \nu_1 - \nu_3)
 (1 + \nu_1 + \nu_3 - \nu_1^2 + \nu_1\nu_3 - \nu_3^2) z^3
\notag \\
&& \mspace{91mu}
+ (2 + \nu_1 - 2\nu_3 - 3\nu_1^2 + 6\nu_1\nu_3
+ \nu_1^3 - 3\nu_1^2\nu_3 - 3\nu_1\nu_3^2 + \nu_3^3) z^4
\notag \\
&& \mspace{91mu}
- 2\nu_3 (1 - \nu_1 + \nu_3) (1 - \nu_1 - \nu_3) z^5
\notag \\
&& \mspace{91mu}
- \nu_3^2 (1 - \nu_1 - \nu_3) z^6
\Biggr],
\label{FV}
\end{eqnarray} 
respectively. 
Here, we define 
the mass ratio as $\nu_I \equiv M_I/M$ 
for the total mass $M \equiv \sum_I M_I$ 
and make a frequent use of $\nu_2=1-\nu_1-\nu_3$. 
It should be noted that 
in this truncated calculation we ignore the second post-Newtonian 
(or higher order) contributions so that 
we can replace, for instance,  
$v_1$ by $R_1 \omega$ (using the Newtonian $R_1$) 
in post-Newtonian velocity-dependent terms such as $v_1^2$. 

In a similar manner to the above Newtonian formulation, 
straightforward but lengthy calculations lead to 
a seventh-order equation as \cite{YA} 
\begin{equation}
F(z) \equiv \sum_{k=0}^7 A_k z^k = 0 , 
\label{7th}
\end{equation}
where we define 
\begin{eqnarray}
A_7 &=& \frac Ma 
\Biggl[
- 4 - 2(\nu_1 - 4\nu_3)
+ 2(\nu_1^2 + 2\nu_1\nu_3 - 2\nu_3^2)
- 2\nu_1\nu_3(\nu_1 + \nu_3)
\Biggr],
\label{A7}
\\
A_6 &=& 1 - \nu_3
+ \frac Ma
\Biggl[
- 13 - (10\nu_1 - 17\nu_3)
+ 2(2\nu_1^2 + 8\nu_1\nu_3 - \nu_3^2)
\notag \\
&& \mspace{103mu}
+ 2(\nu_1^3 - 2\nu_1^2\nu_3 - 3\nu_1\nu_3^2 - \nu_3^3)
\Biggr] ,
\label{A6}
\\
A_5 &=& 2 + \nu_1 - 2\nu_3
+ \frac Ma 
\Biggl[
- 15 -(18\nu_1-5\nu_3) + 4(5\nu_1\nu_3 + 4\nu_3^2)
  \notag \\
&& \mspace{148mu}
+ 6(\nu_1^3 - \nu_1\nu_3^2 - \nu_3^3)
\Biggr] ,
\label{A5}
\\
A_4 &=& 1 + 2\nu_1 - \nu_3
+ \frac Ma
\Biggl[
- 6 - 2(5\nu_1 + 2\nu_3)
- 4(2\nu_1^2 - \nu_1\nu_3 - 4\nu_3^2)
\notag \\
&& \mspace{148mu}
+ 2(3\nu_1^3 + \nu_1^2\nu_3 - 2\nu_1\nu_3^2 - 3\nu_3^3)
\Biggr] ,
\label{A4}
\\
A_3 &=& - (1 - \nu_1 + 2\nu_3)
+ \frac Ma 
\Biggl[
6 + 2(2\nu_1 + 5\nu_3)
- 4( 4\nu_1^2 + \nu_1\nu_3 - 2\nu_3^2)
\notag \\
&& \mspace{177mu}
+ 2( 3\nu_1^3 + 2\nu_1^2\nu_3 - \nu_1\nu_3^2 - 3\nu_3^3)
\Biggr] ,
\label{A3}
\\
A_2 &=& - (2 - 2\nu_1 + \nu_3)
+ \frac Ma 
\Biggl[
15 - ( 5\nu_1 - 18\nu_3)
- 4(4\nu_1^2 + 5\nu_1\nu_3)
\notag \\
&& \mspace{177mu}
+ 6( \nu_1^3 + \nu_1^2\nu_3 - \nu_3^3)
\Biggr] ,
\label{A2}
\\
A_1 &=& - (1 - \nu_1)
+ \frac Ma 
\Biggl[
 13 - ( 17\nu_1 - 10\nu_3)
+ 2( \nu_1^2 - 8\nu_1\nu_3 - 2\nu_3^2)
\notag \\
&& \mspace{130mu}
+ 2( \nu_1^3 + 3\nu_1^2\nu_3 + 2\nu_1\nu_3^2 - \nu_3^3)
\Biggr] ,
\label{A1}
\\
A_0 &=& \frac Ma 
\Biggl[
4 - 2( 4\nu_1 - \nu_3)
+ 2(2\nu_1^2 - 2\nu_1\nu_3 - \nu_3^2)
+ 2\nu_1\nu_3(\nu_1 + \nu_3) 
\Biggr] .
\label{A0}
\end{eqnarray}
Here, the sign of Eq. ($\ref{A0}$) is chosen 
so that it can agree with the fifth-order equation Eq. ($\ref{5th}$) 
in the Newtonian limit of $M/a \to 0$. 
This seventh-order equation is antisymmetric for exchanges 
between $\nu_1$ and $\nu_3$, 
only if one makes a change as $z \to 1/z$. 
This antisymmetry may validate the complicated form of 
each coefficient.

Once a positive root for Eq. ($\ref{7th}$) is found, 
the root $z$ can be substituted into Eq. ($\ref{EOM-PN}$) 
in order to obtain the angular velocity $\omega$. 

The angular velocity including the post-Newtonian effects 
is obtained from Eq. ($\ref{EOM-PN}$) as \cite{YA}
\begin{equation}
\omega = \omega_N 
\left( 1 + \frac{F_M}{2F_N} + \frac{F_V}{2R_{13}} \right) , 
\label{omega}
\end{equation}
where $\omega_N \equiv (F_N/R_{13})^{1/2}$ 
denotes the angular velocity of the Newtonian collinear orbit. 
Note that the slow motion is assumed to derive Eq. (\ref{omega}) 
which is analogous to Kepler's third law.

\section{Existence of positive roots} 
In this section, we show that there always exist  
positive roots for the seventh-order equation 
that has been derived as Eq. ($\ref{7th}$). 
This is nothing but the existence of the post-Newtonian 
collinear solution. 

For later convenience, we recover $\nu_2$ 
and thus rewrite a coefficient $A_0$ as 
\begin{equation}
A_0 = 2\frac Ma (\nu_2 + \nu_3) (2\nu_2 + 2\nu_3 + \nu_2\nu_3) , 
\end{equation}
which immediately leads to 
$A_0 > 0$.  

In a similar manner,  
one can show $A_7 < 0$. 
An alternative but powerful way to see this is using 
the antisymmetry of the seventh-order equation 
for transformations between masses $M_1$ and $M_3$ 
as $\nu_1 \leftrightarrow \nu_3$ 
and $z \leftrightarrow 1/z$. 
This transformation makes a change as 
$A_0 \to -A_7$. 
By using $A_0 > 0$, therefore, 
we have always $A_7 < 0$. 

Bringing the above results together, 
we have $F(0) = A_0 > 0$ and 
$F(\infty) = A_7 z^7|_{z \to \infty} < 0$. 
Therefore, the number of positive roots for $F(z)=0$ 
either equals to one or 
more than it by a multiple of two. 

Let us investigate the seventh-order equation 
in order to more precisely determine the number of positive roots. 
We decompose each coefficient $A_k$ into the Newtonian part $A_{Nk}$ 
and the post-Newtonian one $A_{PNk}$. 
Note that $A_{Nk}$ agrees with the coefficient of $z^{k-1}$ 
(but not $z^k$) in Eq. (\ref{5th}). 
In the Newtonian fifth-order equation by Eq. ($\ref{5th}$), 
we have $A_{N6} > 0$, $A_{N5} > 0$, 
$A_{N4} > 0$, $A_{N3} < 0$, $A_{N2} < 0$, $A_{N1} < 0$. 
In the post-Newtonian approximation, 
the post-Newtonian parts must be much smaller than 
the Newtonian ones 
($|A_{PNk}| \ll |A_{Nk}|$ for each $k$), 
so that the post-Newtonian correction 
cannot change the sign of each coefficient $A_k$. 
We thus have 
$A_{6} > 0$, $A_{5} > 0$, 
$A_{4} > 0$, $A_{3} < 0$, $A_{2} < 0$, $A_{1} < 0$. 
By combining them with $A_7 < 0$ and $A_0 > 0$, 
the number of sign changes of the coefficients 
in Eq. ($\ref{7th}$) is necessarily three. 
Therefore, Descartes' rule of signs indicates that 
Eq. ($\ref{7th}$) has either one or three roots. 
We can easily understand that one of them 
is a correction to the Newtonian orbit. 
What are the other two roots? 
We shall investigate them in next section.

\section{Uniqueness of the post-Newtonian collinear solution} 
\begin{figure}[t]
\includegraphics[width=8cm]{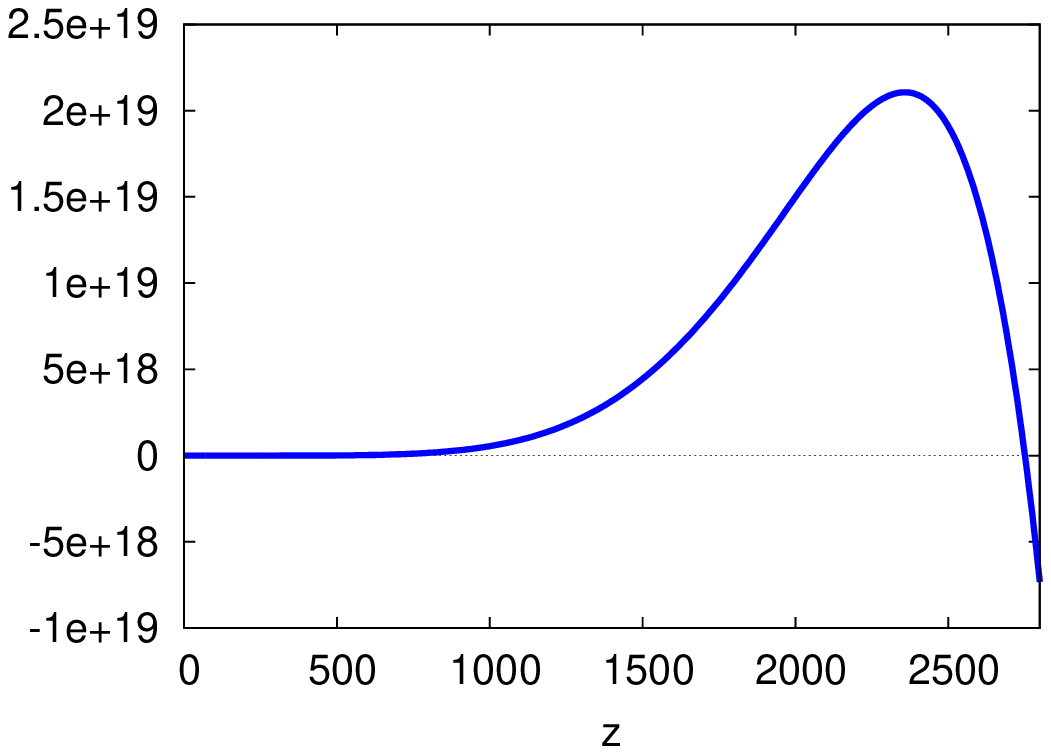}\\
\includegraphics[width=8cm]{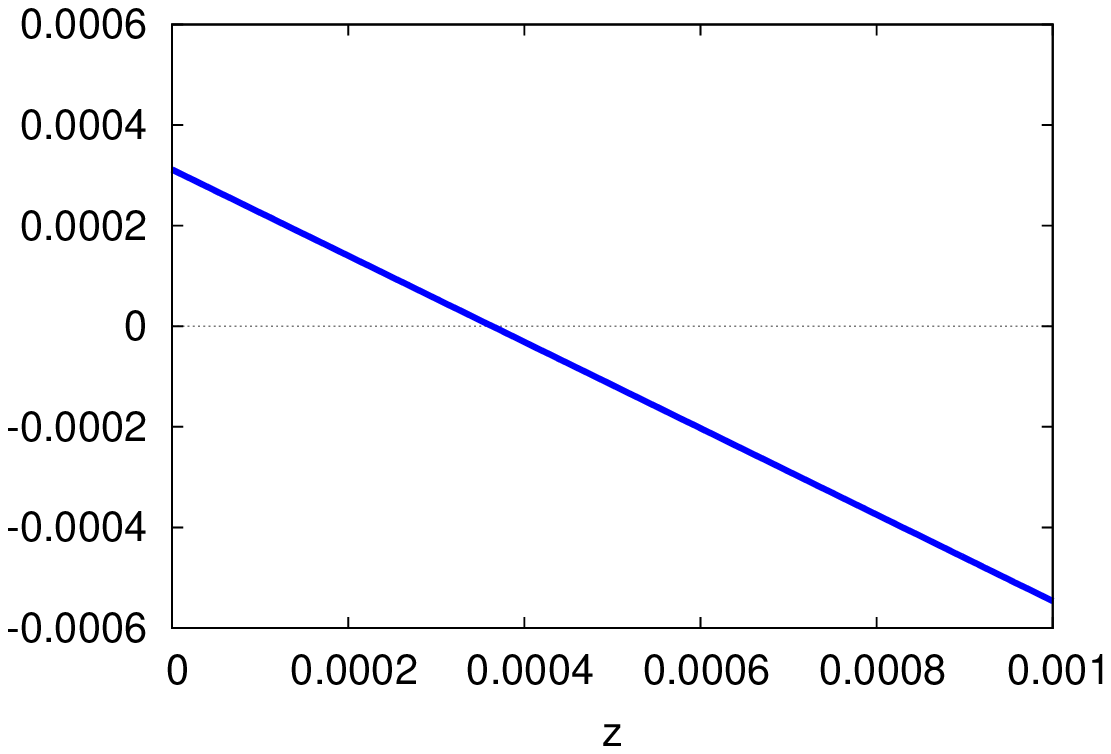}\\
\includegraphics[width=8cm]{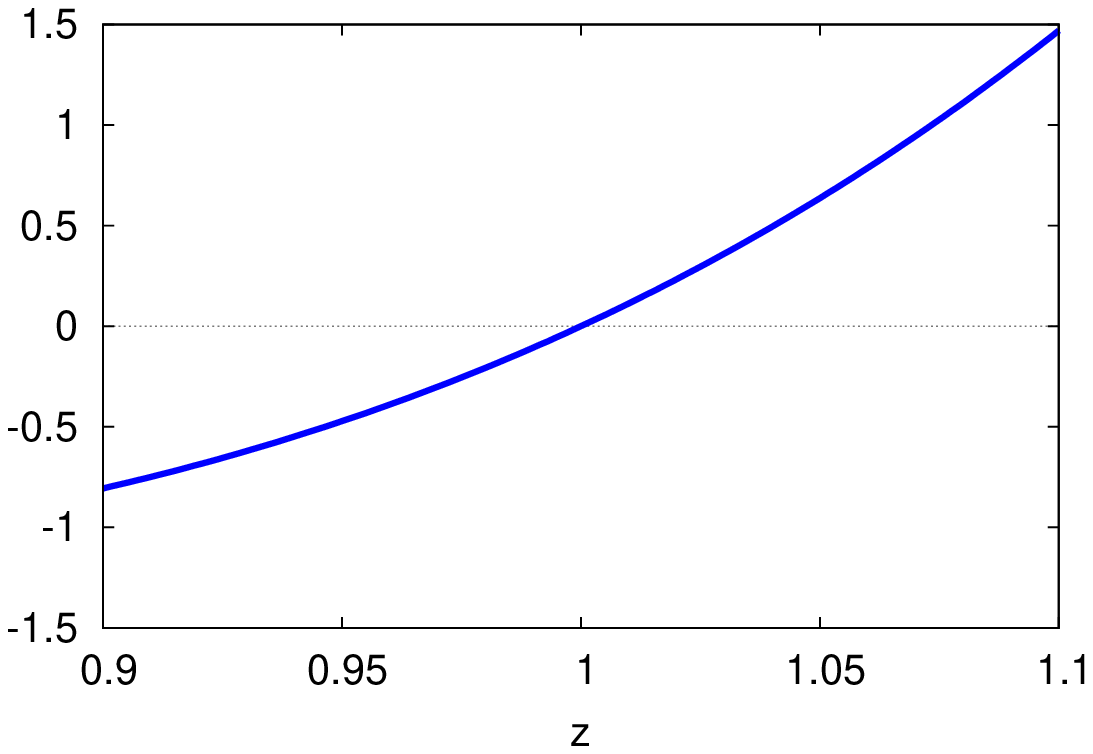}
\caption{ 
Top panel:  
The seventh-order polynomial in the L.H.S. of Eq. ($\ref{7th}$). 
The horizontal axis is chosen as $z$. 
We take $\nu_1 = 1/7$, $\nu_2=5/7$, $\nu_3 = 1/7$, 
$a/M = 10^4$ ($v \sim 10^{-2}$) 
in order to exaggerate small effects in these figures. 
Clearly such a symmetric choice of the mass ratios 
produces a trivial root as $z=1$, 
which makes it easy to check numerical calculations. 
$M_2$ is relatively large so that the centrifugal force can be large. \\
Middle panel: 
The seventh-order polynomial around the smallest positive root $z_S$. \\
Bottom panel: 
The polynomial around the moderate positive root. 
}
\label{f1}
\end{figure}

Figure $\ref{f1}$ shows that the equation has three positive roots, 
where we assume $\nu_1 = 1/7$, $\nu_2=5/7$, $\nu_3 = 1/7$, 
$a/M = 10^4$ ($v \sim 10^{-2}$). 
Table $\ref{table1}$ shows numerical values of 
$z$, $\omega$ and $a\omega$ for Figure $\ref{f1}$. 
Two out of the three roots do not satisfy 
a slow-motion condition for the post-Newtonian approximation 
as shown below.

\begin{table}
\caption{
Values of $z$, $\omega$ and $a\omega$ for Figure $\ref{f1}$. 
Here are three positive roots, where we assume $z_1 < z_2 < z_3$. 
}
  \begin{center}
    \begin{tabular}{llll}
\hline
 & $z_1$ & $z_2$ & $z_3$ \\
\hline 
%\hline
z & 3.635$\times 10^{-4}$ & 1.000 & 2751 \\
\hline 
$\omega$ & 8.723$\times 10^{-5}$ \quad 
& 2.449$\times 10^{-6}$ \quad 
& 8.723$\times 10^{-5}$ \quad  \\
\hline 
$a\omega$ \quad & 0.8723 & 0.02449 & 0.8723 \\
    \end{tabular}
  \end{center}
\label{table1}
\end{table}

Here we show that the remaining two positive roots 
must be discarded. 
Because of the antisymmetry of Eq. ($\ref{7th}$) 
for the transformation as $z \leftrightarrow 1/z$, 
the two roots must be a pair 
through this transformation associated with 
exchanges between $M_1$ and $M_3$. 
Let the smaller root and the larger one be denoted as 
$z_S$ and $z_L$, respectively. 

First, we consider the smallest positive root $z_S$, 
where we assume $z_S \ll 1$. 
Then, Eq.($\ref{7th}$) is approximated as 
\begin{equation}
A_1 z_S + A_0 = 0 , 
\end{equation}
where $A_0$ starts at the post-Newtonian order 
without Newtonian terms 
and $A_1 = A_{N1}+A_{PN1}$ has both the Newtonian terms 
and post-Newtonian corrections ($|A_{N1}| \gg |A_{PN1}|$). 
We thus obtain an approximate form of the smallest root as 
\begin{eqnarray}
z_S &=& - \frac{A_0}{A_{N1}} 
\nonumber\\
&=& O\left( \frac{M}{a} \right) , 
\end{eqnarray}
where we used Eqs. ($\ref{A1}$) and ($\ref{A0}$). 
This implies that 
$z_S$ is indeed of the post-Newtonian order, 
in consistent with $z_S \ll 1$. 
At this point, however, 
we cannot discard this smallest root $z_S$. 

As a next step, let us make an order-of-magnitude estimation 
for the angular velocity $\omega_S$ that satisfies 
Eq. ($\ref{EOM-PN}$) for $z_S$,  
where $\omega_S$ denotes the angular velocity 
corresponding to $z_S$.  
We obtain from Eqs. ($\ref{FN}$), ($\ref{FM}$) and ($\ref{FV}$)
\begin{eqnarray}
F_N &=& O\left( \frac{M}{a^2 z_S^2}\right) 
\nonumber\\
&=& O\left( \frac{1}{M} \right) ,
\label{FN-expS}
\end{eqnarray}
\begin{eqnarray}
F_M &=& O\left( \frac{M^2}{a^3 z_S^3}\right) 
\nonumber\\
&=& O\left( \frac{1}{M} \right) ,
\label{FM-expS}
\end{eqnarray}
\begin{eqnarray}
F_V &=& O\left( \frac{M}{z_S^2}\right) 
\nonumber\\
&=& O\left( \frac{a^2}{M} \right) .
\label{FV-expS}
\end{eqnarray}
We thus find 
$R_{13}=a \ll F_V$ because $M \ll a$. 
Therefore, we find 
$F_N \sim F_M \sim F_V \omega_S^2 \gg R_{13} \omega_S^2$
in Eq. ($\ref{EOM-PN}$). 
This leads to 
\begin{eqnarray}
\omega_S &=& O\left( \frac{1}{a}  \right) ,    
\label{omegaS-exp}
\end{eqnarray}
though $\omega_N^2 = O(M/a^3)$ for the Newtonian case. 
Eq. ($\ref{omegaS-exp}$) implies an extremely fast rotation, 
since the rotational velocity becomes 
$v_S \approx a \omega_S = O(1) $, 
namely, comparable to the speed of light. 
This unacceptable branch of such an extremely fast motion 
contradicts the post-Newtonian approximation 
and does not satisfy Eq. (\ref{omega}). 
Hence, $z_S$ must be abandoned. 

Next, we consider the largest positive root $z_L$, 
where we assume $z_L \gg 1$. 
Then, Eq.($\ref{7th}$) is approximated as 
\begin{equation}
A_7 z_L + A_6 = 0 , 
\end{equation}
where $A_7$ starts at the post-Newtonian order 
without Newtonian terms 
and $A_6 = A_{N6}+A_{PN6}$ has both the Newtonian terms 
and post-Newtonian corrections ($|A_{N6}| \gg |A_{PN6}|$). 
We thus obtain an approximate form of the largest root as 
\begin{eqnarray}
z_L &=& - \frac{A_7}{A_{N6}} 
\nonumber\\
&=& O\left( \frac{a}{M} \right) , 
\end{eqnarray}
where we used Eqs. ($\ref{A7}$) and ($\ref{A6}$). 
This implies that 
$z_L^{-1}$ is indeed of the post-Newtonian order, 
in consistent with $z_L \gg 1$. 
At this point, however, 
we cannot discard this largest root $z_L$. 

As a next step, let us make an order-of-magnitude estimation 
for the angular velocity $\omega_L$ that satisfies 
Eq. ($\ref{EOM-PN}$) for $z_L$, 
where $\omega_L$ denotes the angular velocity 
corresponding to $z_L$. 
We obtain from Eqs. ($\ref{FN}$), ($\ref{FM}$) and ($\ref{FV}$)
\begin{eqnarray}
F_N &=& O\left( \frac{M z_L^2}{a^2}\right) 
\nonumber\\
&=& O\left( \frac{1}{M} \right) ,
\label{FN-expL}
\end{eqnarray}
\begin{eqnarray}
F_M &=& O\left( \frac{M^2 z_L^3}{a^3}\right) 
\nonumber\\
&=& O\left( \frac{1}{M} \right) ,
\label{FM-expL}
\end{eqnarray}
\begin{eqnarray}
F_V &=& O\left( M z_L^2 \right) 
\nonumber\\
&=& O\left( \frac{a^2}{M} \right) .
\label{FV-expL}
\end{eqnarray}
We thus find 
$R_{13}=a \ll F_V$ because $M \ll a$. 
Therefore, we find 
$F_N \sim F_M \sim F_V \omega_L^2 \gg R_{13} \omega_L^2$
in Eq. ($\ref{EOM-PN}$). 
This leads to 
\begin{eqnarray}
\omega_L &=& O\left( \frac{1}{a}  \right) ,    
\label{omegaL-exp}
\end{eqnarray}
though $\omega_N^2 = O(M/a^3)$ for the Newtonian case. 
Eq. ($\ref{omegaL-exp}$) implies an extremely fast rotation, 
since the rotational velocity becomes 
$v_L \approx a \omega_L = O(1) $, 
namely, comparable to the speed of light. 
This unacceptable branch of such an extremely fast motion 
contradicts the post-Newtonian approximation 
and does not satisfy Eq. (\ref{omega}). 
Hence, also $z_L$ must be abandoned. 

We should remember the transformation  
as $z \leftrightarrow 1/z$, namely $1 \leftrightarrow 3$.  
Hence, $z_S$ and $z_L$ correspond to each other 
as $z_S = 1/z_L$.  
In this sense, it seems natural that 
the above argument for discarding $z_L$ 
is very similar to that of $z_S$. 

As a result, two of the three positive roots 
are discarded as unphysical ones. 
Hence, we complete the proof of the uniqueness. 

We mention an application of the uniqueness theorem 
for the restricted three-body problem. 
We have three possibilities for choosing a test mass 
as $M_1=0$, $M_2=0$ or $M_3=0$. 
For each case, we have only the single collinear solution. 
Therefore, the three equilibrium points exist 
along the symmetry axis of the system, 
and they are a generalization of Lagrange points 
$L_1$, $L_2$ and $L_3$. 

Before closing this section, we mention an interesting 
property of the angular velocity of the collinear configurations. 
For the same masses and full length, 
we have always an inequality as 
\begin{equation}
\omega < \omega_N , 
\end{equation}
which means that the post-Newtonian orbital period measured 
in the coordinate time is longer than the Newtonian one. 
Provided that the masses and angular rate are fixed, 
the relativistic length $a$ is shorter than the Newtonian one. 
Detailed calculations are given in the Appendix. 

\section{Conclusion}
%\noindent \emph{Conclusion.--- }
We proved the uniqueness of the collinear configuration 
for given system parameters (the masses and 
the end-to-end length). 
It was shown that the equation 
determining the distance ratio among the three masses, 
which has been obtained as a seventh-order polynomial 
in the previous paper, 
has at most three positive roots, 
which apparently provide three cases of the distance ratio. 
It was found, however, that there exists 
one physically acceptable root and only one. 
The remaining two positive roots are discarded 
in the sense that they do not satisfy the slow motion 
ansatz in the post-Newtonian approximation. 

Especially for the restricted three-body problem, 
exactly three positions of a third body are true even 
at the post-Newtonian order. 
They are relativistic counterparts of 
the Newtonian Lagrange points $L_1$, $L_2$ and $L_3$. 

It was shown also that, for the same masses and full length, 
the angular velocity of the post-Newtonian 
collinear configuration is smaller than that for the Newtonian case.  
Provided that the masses and angular rate are fixed, 
the relativistic length $a$ is shorter than the Newtonian one. 

Our way of discussion seems to work at the second (and higher) 
post-Newtonian orders, because the slow motion approximation 
is a key in the above proof. 
Therefore, the uniqueness of collinear configurations 
for a three-body system may be true even at higher orders, 
precisely speaking, if the configuration has the Newtonian limit. 
It is an open question whether {\it fully} general relativistic 
systems admit a particular solution that can appear 
only for a fast motion case and thus has no Newtonian limit.

We would like to thank the referee for useful comments 
on the earlier version of the manuscript. 
We are grateful to Y. Kojima  
for useful conversations. 
This work was supported in part (H.A.) 
by a Japanese Grant-in-Aid 
for Scientific Research from the Ministry of Education, 
No. 21540252. 

\appendix
\section{Detailed calculations on the angular velocity}
Let us prove $\omega < \omega_N$. 
Eq. ($\ref{omega}$) is rewritten as 
\begin{equation}
\frac{\omega - \omega_N}{\omega_N} 
= 
\frac{F_M R_{13} + F_V F_N}{2 F_N R_{13}} . 
\label{omega-difference}
\end{equation}
Here, $F_N$ is positive and thus the denominator 
of the R.H.S. of Eq. ($\ref{omega-difference}$) is always positive. 
What we have to do is to investigate the sign of the numerator 
for the R.H.S. of Eq. ($\ref{omega-difference}$). 
The numerator is factored as 
\begin{equation}
\frac{M^2}{a^2 z^4 (1+z)^2} 
\times 
\sum_{k=0}^{10} a_k z^k , 
\label{numerator}
\end{equation}
%where each coefficient $a_k$ is defined as ...
where we define
\begin{eqnarray}
a_{10} &=& - (1 - \nu_1 - \nu_3)^2 \nu_3^2 , 
\label{a10}\\
\notag\\
a_9 &=& - (1 - \nu_1 - \nu_3) 
(4 + \nu_1 - 2\nu_3 - 4\nu_1\nu_3 + 2\nu_3^2 + 2\nu_1^2\nu_3 -
2\nu_1\nu_3^2 - 4\nu_3^3) , 
\label{a9}\\
\notag\\
a_8 &=& - (1 - \nu_1 - \nu_3)
(18 + 4\nu_1 - 9\nu_3 + 3\nu_1^2 - 14\nu_1\nu_3 \notag\\
&& \mspace{120mu}
 + 2\nu_3^2 - \nu_1^3 + 7\nu_1^2\nu_3 + 2\nu_1\nu_3^2 - 6\nu_3^3) , 
\label{a8}\\
\notag\\
a_7 &=& - (1 - \nu_1 - \nu_3)
(32 + 4\nu_1 - 13\nu_3 + 12\nu_1^2 - 18\nu_1\nu_3 \notag\\
&& \mspace{120mu}
+ 10\nu_3^2 - 4\nu_1^3 + 8\nu_1^2\nu_3 + 4\nu_1\nu_3^2 - 8\nu_3^3), 
\label{a7}\\
\notag\\
a_6 &=& - (30 - 30\nu_1 - 37\nu_3 + 19\nu_1^2 - 12\nu_1\nu_3 + 27\nu_3^2
 - 22\nu_1^3 + 18\nu_1^2\nu_3 \notag\\
&& \mspace{20mu}
+ 12\nu_1\nu_3^2 - 28\nu_3^3 + 6\nu_1^4 - 4\nu_1^3\nu_3 - 15\nu_1^2\nu_3^2 + 6\nu_1\nu_3^3 +
11\nu_3^4) , 
\label{a6}\\
\notag\\
a_5 &=& 
- 2 (12 - 13\nu_1 - 13\nu_3 + 11\nu_1^2 - 10\nu_1\nu_3 + 11\nu_3^2
 - 11\nu_1^3 + 17\nu_1^2\nu_3 \notag\\
&& \mspace{30mu}
+ 17\nu_1\nu_3^2 - 11\nu_3^3 + 4\nu_1^4 - 3\nu_1^3\nu_3 - 14\nu_1^2\nu_3^2 
- 3\nu_1\nu_3^3 + 4\nu_3^4) , 
\label{a5}\\
\notag \\
a_4 &=&
- (30 - 37\nu_1 - 30\nu_3 + 27\nu_1^2 - 12\nu_1\nu_3 + 19\nu_3^2
 - 28\nu_1^3 + 12\nu_1^2\nu_3 \notag\\ 
&& \mspace{20mu}
+ 18\nu_1\nu_3^2 - 22\nu_3^3
 + 11\nu_1^4 + 6\nu_1^3\nu_3 - 15\nu_1^2\nu_3^2 - 4\nu_1\nu_3^3 +
 6\nu_3^4) , 
\label{a4}\\
\notag\\
a_3 &=& - (1 - \nu_1 - \nu_3)
(32 - 13\nu_1 + 4\nu_3 + 10\nu_1^2 - 18\nu_1\nu_3 \notag\\
&& \mspace{120mu}
+ 12\nu_3^2 - 8\nu_1^3 + 4\nu_1^2\nu_3 + 8\nu_1\nu_3^2 - 4\nu_3^3) , 
\label{a3}\\
\notag\\
a_2 &=& - (1 - \nu_1 - \nu_3)
(18 - 9\nu_1 + 4\nu_3 + 2\nu_1^2 - 14\nu_1\nu_3 \notag\\
&& \mspace{120mu}
+ 3\nu_3^2 - 6\nu_1^3 + 2\nu_1^2\nu_3 + 7\nu_1\nu_3^2 - \nu_3^3) , 
\label{a2}\\
\notag\\
a_1 &=& - (1 - \nu_1 - \nu_3)
(4 - 2\nu_1 + \nu_3 + 2\nu_1^2 - 4\nu_1\nu_3
 - 4\nu_1^3 - 2\nu_1^2\nu_3 + 2\nu_1\nu_3^2) , 
\label{a1}\\
\notag\\
a_0 &=&  - (1 - \nu_1 - \nu_3)^2\nu_1^2 .\label{a0}
\end{eqnarray}

We show $a_k<0$ for each $k$.
It is trivial that $a_0<0$ and $a_{10}<0$.
For $\nu_1 \leftrightarrow \nu_3$, we have a symmetry between
$a_9 \leftrightarrow a_1$, $a_8 \leftrightarrow a_2$,
$a_7 \leftrightarrow a_3$ and $a_6 \leftrightarrow a_4$.
Therefore, it is sufficient to examine $a_9$, $a_8$, $a_7$, $a_6$ and
$a_5$.

First, let us show $a_9<0$.
The nontrivial factor in Eq. (\ref{a9}) turns out to be positive by
noting the following relation as
\begin{eqnarray}
&&
4 - 2\nu_3 - 4\nu_1\nu_3 + 2\nu_3^2 - 2\nu_1\nu_3^2 - 4\nu_3^3 \notag\\
&=&
4(1 - \nu_3)^3 + 10\nu_3(1 - \frac4{10}\nu_1 - \nu_3 -  
\frac2{10}\nu_1\nu_3) \notag\\
&>&
4(1 - \nu_3)^3 + 10\nu_3(1 - \frac6{10}\nu_1 - \nu_3) \notag\\
&>&
4(1 - \nu_3)^3 + 10\nu_3(1 - \nu_1 - \nu_3) \notag\\
&>& 0 ,
\end{eqnarray}
where we used $\nu_1 \geqq 0$, $\nu_3 \leqq 1$, $\nu_1\nu_3 \leqq \nu_1$.
Hence we find $a_9<0$.

We discuss the sign of $a_8$.
By using $\nu_1 = 1 - \nu_2 - \nu_3$ to delete $\nu_1$ and recover
$\nu_2$, the R.H.S. of Eq. (\ref{a8}) is factored as
\begin{eqnarray}
- \nu_2 ( 24 - 7\nu_2 - 23\nu_3 + 4\nu_3^2 + \nu_2^3 
+ 10\nu_2^2\nu_3 + 15\nu_2\nu_3^2) .
\label{a8-1}
\end{eqnarray}
One can show that the latter three terms are positive, since 
\begin{eqnarray}
24 - 7\nu_2 - 23\nu_3 
&=& 
1 + 7(1 - \nu_2 - \nu_3) + 16(1 - \nu_3)\notag\\
&>& 0 .
\end{eqnarray}
Hence, the second factor in Eq. (\ref{a8-1}) is always positive, which
leads to $a_8<0$, and also $a_2<0$.

Next, we examine $a_7$.
By recovering $\nu_2$ to delete $\nu_1$, 
the R.H.S. of Eq. (\ref{a7}) is factored as
\begin{eqnarray}
- \nu_2 ( 44 - 16\nu_2 - 39\nu_3 + 2\nu_2\nu_3 + 16\nu_3^2 
+ 4\nu_2^3 + 20\nu_2^2\nu_3 + 24\nu_2\nu_3^2) .
\end{eqnarray}
A key thing is a positive as
\begin{eqnarray}
44 - 16\nu_2 - 39\nu_3 
&=& 
5 + 16 (1 - \nu_2 - \nu_3) + 23(1 - \nu_3) \notag\\
&>& 0 ,
\end{eqnarray}
which immediately leads to $a_7<0$, and also $a_3<0$.

We investigate $a_6$.
Similarly to $a_7$, it is factored as
\begin{eqnarray}
- ( 3 + 34\nu_2 - \nu_3 - 11\nu_2^2 - 34\nu_2\nu_3 + \nu_3^2 
- 2\nu_2^3 + 24\nu_2\nu_3^2 + 6\nu_2^4 + 28\nu_2^3\nu_3 + 33\nu_2^2\nu_3^2 ).
%\notag\\
\label{a6-1}
\end{eqnarray}
Here, we see that the following two combinations both are positive,
\begin{eqnarray}
34\nu_2 - 11\nu_2^2 - 34\nu_2\nu_3 
&=&
11\nu_2(1 - \nu_2 - \nu_3) + 23\nu_2(1 - \nu_3) \notag\\
&>& 0,\\
3 - \nu_3 - 2\nu_2^3
&>& 
2 - 2\nu_2^3 \notag\\
&>& 0,
\end{eqnarray}
which show that Eq. (\ref{a6-1}) is always negative.
Hence, we show $a_6<0$, and also $a_4<0$.

Also for $a_5$, it is factored as
\begin{eqnarray}
&&- 2 ( 3 + 8\nu_2 - \nu_3 + 2\nu_2^2 - 11\nu_2\nu_3 + \nu_3^2
- 5\nu_2^3 \notag\\
&&\mspace{30mu}
- 7\nu_2^2\nu_3 + 12\nu_2\nu_3^2 + 4\nu_2^4 
+ 19\nu_2^3\nu_3 + 19\nu_2^2\nu_3^2 ).
\end{eqnarray}
One can find the following rather tricky manipulation as
\begin{eqnarray}
&&
3 + 8\nu_2 - \nu_3 + 2\nu_2^2 - 11\nu_2\nu_3 - 5\nu_2^3 - 7\nu_2^2\nu_3
\notag\\
&=& 
3 + 5\nu_2[1 - \nu_2(\nu_2 + \nu_3) - \nu_3] - \nu_3 - 6\nu_2\nu_3
\notag\\
&>&
3 + 5\nu_2(1 - \nu_2 - \nu_3) - \nu_3 - 6\nu_2\nu_3
\notag\\
&>& \frac12,
\end{eqnarray}
where we used $0<\nu_2+\nu_3<1$ and $\nu_2\nu_3 \leqq 1/4$.
Hence, we find $a_5<0$.

As a consequence, all the coefficients are always negative,
which shows $\omega < \omega_N$ for any mass ratio.


\begin{thebibliography}{99}
\bibitem{Danby}
J. M. A. Danby, {\it Fundamentals of Celestial Mechanics} 
(William-Bell, VA, 1988). 
\bibitem{Goldstein}
H. Goldstein, {\it Classical Mechanics} 
(Addison-Wesley, MA, 1980). 
\bibitem{THA}
  Y.~Torigoe, K.~Hattori and H.~Asada,
  Phys.\ Rev.\ Lett.\  {\bf 102}, 251101 (2009). 
\bibitem{Asada}
H. Asada, Phys. Rev. D {\bf 80} 064021 (2009).
\bibitem{SM}
N. Seto, T. Muto, Phys. Rev. D {\bf 81} 103004 (2010).
\bibitem{Schnittman}
J. D. Schnittman, arXiv:1006.0182. 
\bibitem{Nordtvedt}
K. Nordtvedt,  
Phys. Rev. {\bf 169} 1014 (1968).
\bibitem{Brumberg}V. A. Brumberg,  
{\it Essential relativistic celestial mechanics}, 
(Bristol, UK: Adam Hilger, 1991). 
\bibitem{GAIA}
http://www.rssd.esa.int/index.php?project=GAIA\&page=index.
\bibitem{JASMINE}
http://www.jasmine-galaxy.org/index.html.
\bibitem{Klioner}
S. A. Klioner, 
Astron. J.  {\bf 125} 1580 (2003). 
\bibitem{ICTN}
K. Ioka, T. Chiba, T. Tanaka, T. Nakamura, 
Phys. Rev. D {\bf 58}, 063003 (1998). 
\bibitem{Wardell}
Z. E. Wardell, 
Mon. Not. R. Astron. Soc. {\bf 334}, 149 (2002). 
\bibitem{CDHL}
M. Campanelli, M. Dettwyler, M. Hannam, C. O. Lousto, 
Phys. Rev. D {\bf 74}, 087503 (2006). 
\bibitem{GMH}
K. Gultekin, M. C. Miller, D. P. Hamilton, 
Astrophys.J. {\bf 640} 156 (2006).
\bibitem{Marchal}
C. Marchal, {\it The Three-Body Problem} 
(Elsevier, Amsterdam, 1990). 
\bibitem{Moore} 
C. Moore, 
Phys. Rev. Lett. {\bf 70}, 3675 (1993). 
\bibitem{CM} 
A. Chenciner, R. Montgomery, 
Ann. Math. {\bf 152}, 881 (2000). 
\bibitem{ICA}
  T.~Imai, T.~Chiba and H.~Asada,
  %``Choreographic solution to the general relativistic three-body problem,''
  Phys.\ Rev.\ Lett.\  {\bf 98}, 201102 (2007). 
\bibitem{LN}
  C.~O.~Lousto and H.~Nakano, 
  Class.\ Quant.\ Grav.\  {\bf 25}, 195019 (2008). 
\bibitem{YA}
K. Yamada, H. Asada, Phys. Rev. D {\bf 82}, 104019 (2010). 
\bibitem{Waerden}B. L. van der Waerden, 
{\it Algebra I} (Springer, Berlin 1966). 
\bibitem{MTW}
C. W. Misner, K. S. Thorne, J. A. Wheeler, 
{\it Gravitation}, 
(Freeman, New York, 1973).
\bibitem{LL}L. D. Landau and E. M. Lifshitz, {\it The Classical Theory 
of Fields} (Oxford, Pergamon 1962). 
\end{thebibliography}
\end{document}